\documentclass[twocolumn]{aa}
\usepackage{graphicx}
\usepackage{subcaption}
\usepackage{txfonts}
\begin{document} 

   \title{An elliptic expansion of the potential field source surface model}
   \titlerunning{Elliptic expansion of the PFSS model}


   \author{M. Kruse                             \inst{1}
          \and V. Heidrich-Meisner              \inst{1}
          \and R.F. Wimmer-Schweingruber        \inst{1}
          \and M.Hauptmann                      \inst{2}
          }

   \institute{University of Kiel, Institute for Experimental and Applied Physics,
              Leibnizstr. 11, 24118 Kiel\\
              \email{kruse@physik.uni-kiel.de}
         \and
             University of Kiel, Department of Mathematics, Ludewig-Meyn-Str. 4, 24118 Kiel\\
             }

   \date{Received ***}

 
  \abstract
   {The potential field source surface model is frequently used as a basis for further scientific investigations where a comprehensive coronal magnetic field is of importance. Its parameters, especially the position and shape of the source surface, are crucial for the interpretation of the state of the interplanetary medium. Improvements have been suggested that introduce one or more additional free parameters to the model, for example, the current sheet source surface (CSSS) model.}
  {Relaxing the spherical constraint of the source surface and allowing it to be elliptical gives modelers the option of deforming it to more accurately match the physical environment of the specific period or location to be analyzed.}
   {A numerical solver is presented that solves Laplace's equation on a three-dimensional grid using finite differences. The solver is capable of working on structured spherical grids that can be deformed to create elliptical source surfaces. }
   {The configurations of the coronal magnetic field are presented using this new solver. Three-dimensional renderings are complemented by Carrington-like synoptic maps of the magnetic configuration at different heights in the solar corona. Differences in the magnetic configuration computed by the spherical and elliptical models are illustrated.}
    {}
    

   \maketitle
%

\section{\label{sec:introduction}Introduction}

The Sun's coronal magnetic field configuration is an important component in understanding the physics of the heliosphere and the solar dynamo. Due to the vast extent of the heliosphere, a direct measurement of its global structure is not possible. Therefore, computational modeling tools are employed to approximate its structure. Modern high-accuracy algorithms, like full magnetohydrodynamic (MHD) solvers, compute a vast set of physical phenomena and require significant computing power. These models rely on additional modeling assumptions that are difficult to verify and on boundary conditions that have to be modeled because they cannot be observed directly.

Simpler models exist that produce less precise results, but that can be computed orders of magnitude more quickly. For some scientific efforts like long-running data evaluation tasks, for example characterizing solar wind streams, rapid computation of the heliospheric magnetic field over long periods is crucial, while the overall accuracy can be lower without affecting the large-scale results of these studies.

One of the earliest models that was used to model the solar coronal magnetic field is the potential field source surface (PFSS) model \citep{Altschuler1969,Schatten1969}. A brief description of this model is presented in Sect. \ref{sec:SHC}.  The PFSS model returns an analytic expression that allows us to  predict the magnetic field configuration of the Sun between the photosphere and a virtual spherical surface, called the source surface, at a height of a few solar radii  in the corona. The boundary condition for the PFSS model is that all magnetic field lines have to be oriented radially at the source surface, which is in accordance with observations farther out in the heliosphere.

An improved version of this model is the current sheet source surface (CSSS) model \citep{Zhao1995a}, which  adds a second virtual sphere, the cusp surface, between the photosphere and source surface. Above this surface, all magnetic field lines are required to be open, but do not have to be oriented radially. This allows  more freedom when modeling the magnetic field.

Slightly more sophisticated models build upon the force-free approach which neglects external forces on a restricted domain near the solar surface. The most general group of these models are the nonlinear force-free models \citep{Aly1989, Wiegelmann2008}. 
Simplifying this physical model by assuming a current-free domain in addition to it being force-free leads to the PFSS model that allows the extrapolation of the coronal magnetic field using only the line-of-sight component of the photospheric magnetic field configuration (also called a synoptic magnetogram). The photospheric magnetograms can be obtained from direct observations and the procedure has been applied for many decades. While the PFSS model itself makes more simplifying assumptions than the full MHD approach, it relies less on modeled parameters, which can potentially introduce false or inaccurate assumptions to the model.

The assumption that the source surface is spherical   simplifies the mathematical framework as well as the computations significantly. Without hints as to   exactly what the source surface looks like, as well as a lack of computing power back when this model was created,  the only reasonable starting point was to assume a spherical source surface. Also, the very low resolution of available magnetic observations of the photosphere at that time made more accurate model assumptions meaningless.

Since then, several suggestions have been made that question the merit of the spherical source surface. Schulz et al. (\citeyear{Schulz1978, Schulz1997}) have developed an algorithm to alter the shape of the source surface. They proposed surfaces of constant magnetic flux (isogauss) to function as the source surface, and presented a model with a surface that has greater heights above the poles compared to the equator. In this model the free parameter of the PFSS model (the source surface height $R_{ss}$) is substituted for the constant magnetic flux $B_0$ of the isogauss surface, which is determined by a hypothetical solar internal dipole and the constraint of magnetic field lines to be normal to this surface. The strategy behind this approach is to better match field lines to the predictions of an MHD implementation that produces field lines that are not quite radially oriented at the height of the spherical source surface.

Expanding on this idea \cite{Levine1982} proposed a nonspherical source surface that is determined by three free parameters, the mean height and two parameters defining the shape. \cite{Levine1982} also deviated from the constraint of  magnetic field lines that are strictly perpendicular to the source surface, and compared the orientation of computed field lines to total solar eclipse observations.

\cite{Riley2006}   computed isosurfaces of $B_r/|\vec{B}|=0.97$ by employing a magnetohydrodynamic solver and found shapes that resemble prolate spheroids with indentations at the poles. While these isosurfaces do not constitute the source surface, they illustrate the deviation of the magnetic field configuration from spherical symmetry.

The spherical source surface is the simplest assumption and it allows fast computation of the solar magnetic field. However, observations show that the physical conditions of the outer solar coronal plasma depend on longitude as well as latitude, and that  polar regions exhibit a more radial orientation of the magnetic field than at the equator \citep[see, e.g., ][]{McComas1998}. An oblate elliptical source surface creates a magnetic field that is consistent with this observation, as is discussed in Sect. \ref{sec:ellgrid}.

In the past decades the increase in computing power and  space-bound magnetic field observations, have given rise to more intricate models. MHD solvers have allowed a wider array of physical phenomena to be incorporated into the derivation of the solar magnetic field. While potentially more accurate, these models require a larger set of input parameters and model assumptions that are difficult to verify with current observations. The simplicity of the PFSS model allows calculating the rough structure of the solar magnetic field comparably quickly. It is therefore desirable to improve this simple model without adding the complexity of a full MHD approach.

In this work we suggest modifying the classical PFSS model to incorporate elliptical source surfaces. This increases the number of free parameters from one (the source surface height $R_{ss}$) to two (adding the ellipticity or deviation from sphericity of the source surface $A$). In contrast to the approaches mentioned above, we chose to implement a finite difference solver rather than an analytical solver. This allows  easier adjustments of the utilized source surface and gives us the possibility to increase solution accuracy by adding more grid points to regions of strong magnetic gradients. We hope that this minor adjustment aids in the modeling of the large-scale structure of the solar magnetic field more accurately and gives insight into the reliability of the PFSS model for different stages of the solar activity cycle.

Evaluating the validity and accuracy of the various existing models is a field of study of its own. Since the phenomena in question cannot be recreated in their entirety in a laboratory environment, the scientific community is forced to browse sparsely available spacecraft data to look for indicators of correctness or lack thereof. In this work we focus on presenting alterations to the classical PFSS model and the computational results produced by them. In Sect. \ref{sec:conclusion} we discuss options of evaluating the results produced by the different PFSS models.

\section{\label{sec:methods}Methods}

The PFSS model is often used to investigate the link between the interplanetary medium and the solar surface. Because of its simplicity it has strengths and weaknesses, which we investigate here by implementing three different versions of the PFSS. In Section \ref{sec:PFSS} we give a brief summary of the mathematical framework underlying all the implementations. The first implementation recreates the framework developed by \cite{Zhao1993} so that subsequent alterations to the model can be compared to a version that has been thoroughly evaluated. This classical implementation is briefly described in Sect. \ref{sec:SHC}. We created an implementation of this approach and call it the spherical harmonic coefficient (SHC) version. The second version is a numerical finite differences solver that solves Laplace's Equation (Equation \ref{equ:gauss}) on several grid points throughout the computational domain of the PFSS model and is described in Sect. \ref{sec:kielgrid}. The third version is an alteration of the second;  it allows for an elliptical source surface and is described in Sect. \ref{sec:ellgrid}. A summary of our treatment of input data as well as the parameters for the computational solver are presented in Sect. \ref{sec:datatreatment}. 

\subsection{\label{sec:PFSS}The PFSS model}
\begin{figure}
\centering
\includegraphics[angle=-90,width=\hsize]{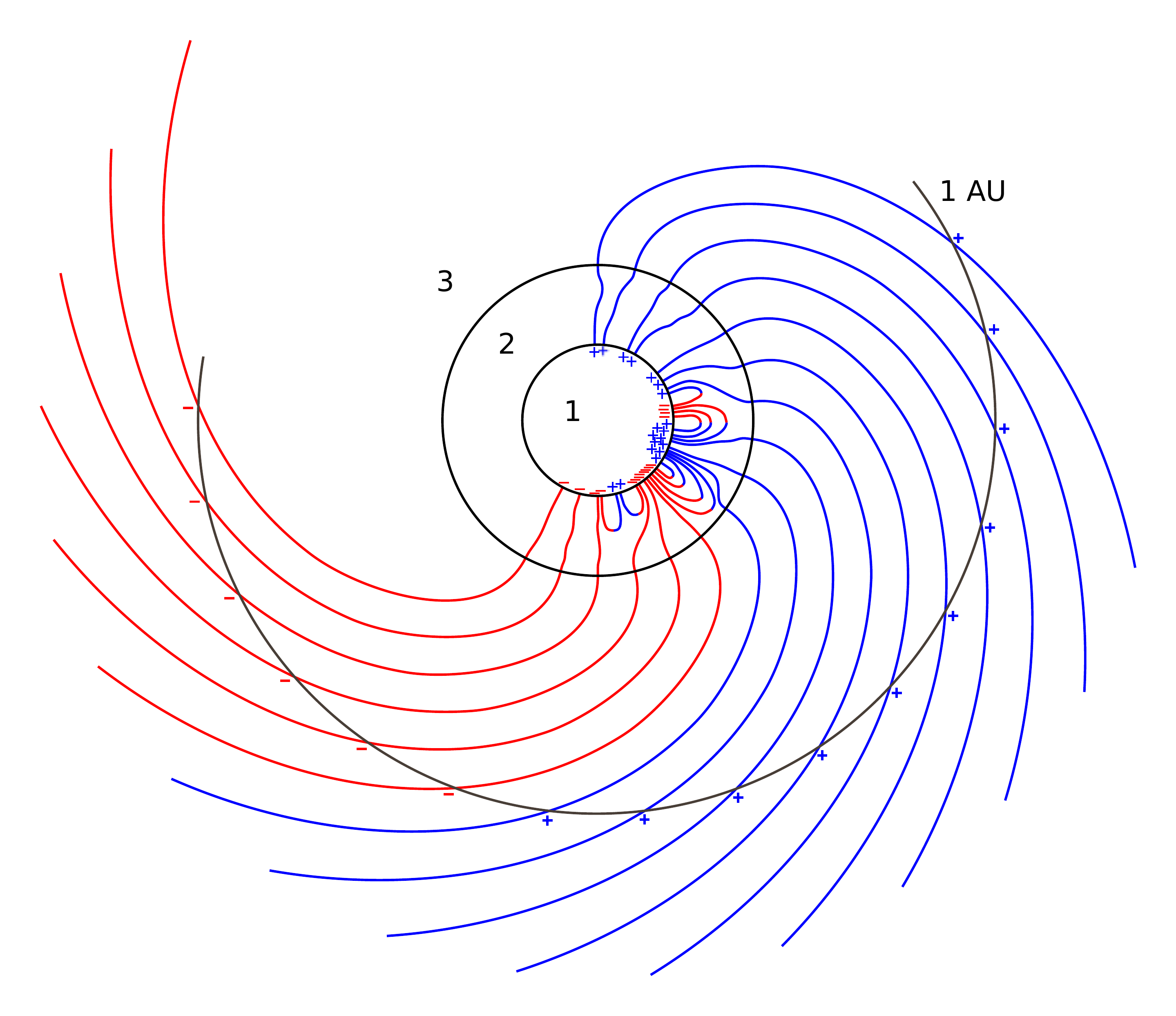}
\caption{Three regions underlying the PFSS model: Region 1 is the inside of the Sun, region 2 is the computational domain between the photosphere and the source surface, and region 3 is the interplanetary space where the solar wind flows radially outward. Figure adapted from \cite{Schatten1969}.\label{fig:PFSSregions}}
\end{figure}

\citet{Altschuler1969} and  \citet{Schatten1969} independently proposed a magnetostatic model for the solar corona. They partition the domain of interest into three regions (Fig. \ref{fig:PFSSregions}). 

Employing several instruments the magnetic field can be measured at the photosphere, which is the boundary separating regions 1 and 2. Region 2 is the computational domain of the PFSS model. The boundary between regions 2 and 3 is called the source surface. Above the source surface, the solar wind dominates the magnetic field configuration (region 3). At least during the quiet times of the solar activity cycle, and aside from violent eruptions, the Sun's photosphere displays features that persist for several Carrington rotations. Therefore, in a first approach the lower region of the solar corona (region 2) can be assumed to be electrostatic, or $\frac{\partial\vec{E}}{\partial t} = \vec{0}$. Furthermore, it is assumed that due to the sharp decrease in particle density above the photosphere and with a smaller decrease in magnetic field strength, the electric current density can be neglected to some point, or $\vec{j}=\vec{0}$. Ampère's law then states 

\begin{equation*}
    \nabla \times \vec{B} = \mu_0\left(\vec{j} + \epsilon_0\frac{\vec{\partial E}}{\partial t}\right) = \vec{0},
\end{equation*}
where $\vec{B}$ is the magnetic flux density, $\vec{E}$ is the electric field, $\vec{j}$ is the electric current density and $\mu_0$ and $\epsilon_0$ are the permeability and permittivity of free space, respectively.

A curl-free vector field can be described as the gradient of a scalar potential ($\nabla\times\nabla f = \vec{0}$ for twice the continuously differentiable $f$, the curl of a gradient vanishes everywhere). Therefore, we write $\vec{B}=-\nabla\Psi$, where $\Psi$ is the scalar magnetic potential. Gauss's law then states

\begin{equation}
\label{equ:gauss}
    \nabla\cdot\vec{B} = -\nabla\cdot\nabla\Psi = -\Delta\Psi = 0.
\end{equation}

Equation \ref{equ:gauss} can be integrated within region 2, given two boundary conditions, one at the photosphere and one at the source surface. At some height the solar wind carries the magnetic field outward, where it is said to be frozen in. The solar wind is advected outward radially, so at the source region of this flow, the magnetic field lines have to be aligned radially as well. Therefore the upper boundary condition is given by the restriction of the magnetic field lines to be perpendicular to the source surface (Neumann boundary condition). For the lower boundary, the photosphere, the magnetic field configuration is known and is supplied to the algorithms by synoptic (line of sight) magnetograms; therefore, a Dirichlet boundary condition is applied. 

For the lower boundary condition, two main approaches are employed regularly by the scientific community, called the radial approach and the line-of-sight approach \citep{Wang1992,Altschuler1969}. Due to its simplicity and the widespread availability of data for comparison, we employ the radial approach for this study.

\subsection{\label{sec:SHC} Spherical harmonic coefficient implementation}

The mathematical framework for the approach to model the solar coronal magnetic field using spherical harmonic coefficients and associated Legendre polynomials can be found in \citet{Altschuler1969} and \citet{Chapman1940}. All that is necessary to recreate the magnetic field is an implementation of the associated Legendre polynomials $P_l^m$ as well as the computed harmonic coefficients $g_{lm}$ and $h_{lm}$ \citep[for explanations of these symbols, see][]{Chapman1940}. The magnetic field configuration at any point in region 2 of the PFSS model can then be acquired by evaluating an analytic expression \citep[see, e.g.,  Eqs. 8--10 in][]{Altschuler1969}.

Today the harmonic coefficients are computed by several groups, including the John M. Wilcox Solar Observatory at Stanford University \citep{Zhao1993} and published on their website \citep{Hoeksema2020}. We used synoptic magnetograms from the Wilcox Solar Observatory \citep{Duvall1977} and compared the results from the Stanford PFSS implementation with our own. Within floating-point precision, our SHC solver produces the same coefficients as  published by Stanford.

\subsection{\label{sec:kielgrid}Grid approach}

Instead of fitting the observed photospheric magnetograms to spherical harmonic functions, we have developed a numeric solver that works on a three-dimensional grid and employs finite differences. This allows us to deform the grid in other implementations to incorporate elliptical source surfaces, while also giving us a tool to compare our results with the implementations of other groups. 

The computational grid stretches from the photosphere (at $r=R_\odot$) to the source surface in the radial direction, from the northern boundary supplied by the magnetogram to the southern boundary in the meridional direction and around the sphere in the zonal direction without boundaries. Grid spacing is equidistant in zonal direction and follows a sine-latitude distribution in meridional direction, as do the underlying synoptic magnetograms. In the radial direction spacing between grid points increases from the photosphere to the source surface geometrically (see Sect. \ref{sec:datatreatment} for more details).

The solution method is an explicit time-stepping algorithm that solves Laplace's equation \ref{equ:gauss} at each grid point for the potential field $\Psi$. This is done by evaluating the analytic expression utilizing finite differences and solving for the magnetic potential $\Psi_{ijk}$, where $i$, $j$, and $k$ denote the radial, meridional, and zonal position in the numerical grid, respectively.

In the initial state, the potential field is set to zero at each grid point except the lowest grid shell which is derived from the synoptic line-of-sight magnetograms. The potential at the uppermost grid shell (at the source surface, $r=R_{ss}$) is kept constant throughout the iteration process, thus implementing the radial boundary condition at the source surface. The potential at each grid point is stored and compared to that obtained in the following time step. Let $\Psi_{ijk}^{t}$ denote the magnetic potential at position $i,j,k$ in time step $t$, and $\Psi_{ijk}^{t-1}$ the potential computed in the previous time step. The algorithm terminates if the maximum relative deviation to the previous time step at all grid points drops below a specified accuracy threshold $p$, or

\begin{equation}
    \max_{ijk}\left(e_{ijk}\right)=\max_{ijk}\left(\frac{|\Psi_{ijk}^{t}-\Psi_{ijk}^{t-1}|}{|\Psi_{ijk}^{t}|}\right) < p.
    \label{equ:termination}
\end{equation}

The numerical grid solver produces nearly the same polarity configuration at the source surface as the SHC implementation. Closer to the photosphere the numerical solver can resolve a finer structure than the SHC solver, which is discussed in section \ref{sec:shccomp}. 

\subsection{\label{sec:ellgrid}Grid approach with an elliptic source surface}

    \begin{figure*}
        \centering
        \includegraphics{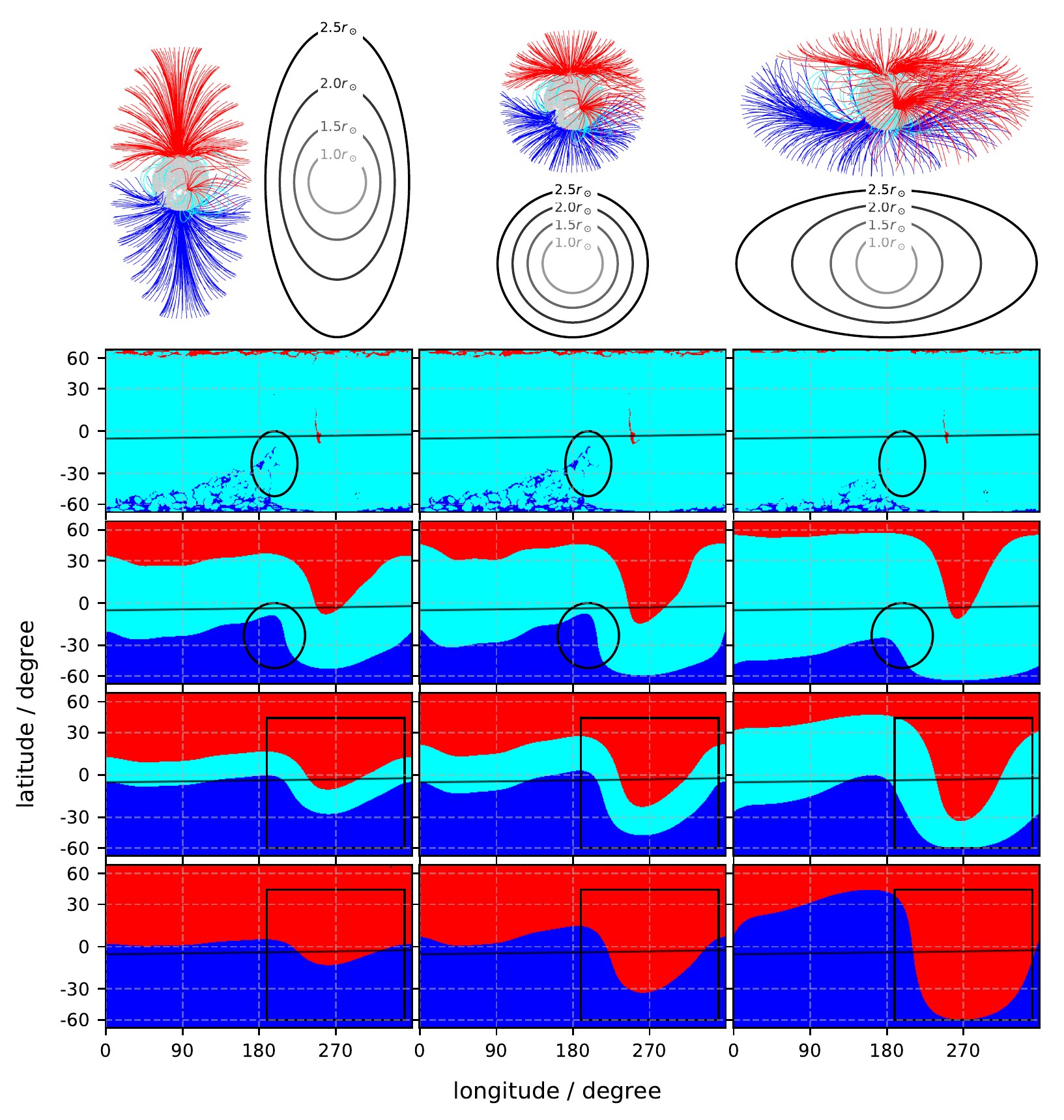}
        \caption{Magnetic field polarity configuration at different heights.   Results shown for a prolate ellipsoidal source surface with ellipticity 2.0 (left column),  for the classical spherical source surface (center), and  for an oblate ellipsoidal surface again with ellipticity 2.0 (right). All figures were created using our grid solver. The source surface height for all models is 2.5 $R_\odot$ (minor half-axis in the ellipsoidal cases). Depicted is Carrington rotation 2066. Data for the lower boundary was obtained from MDI onboard SOHO. The first row shows a three-dimensional rendering of a few magnetic field lines for each model as well as a cut through the height levels depicted below. Rows 2 to 5 show the magnetic field polarity configuration at height levels $1.0 R_\odot$, $1.5R_\odot$, $2.0R_\odot$, and $2.5R_\odot$ (minor half-axis/radius) in a synoptic Carrington format. The height levels correspond to ellipsoids at distances of 0\%, 33\%, 67\%, and 100\% between the photosphere and source surface. Red magnetic field lines and pixels are directed inward, blue lines are directed outward, and cyan indicates closed field structures. The projection of Earth on the height levels is drawn as a black ascending line near the solar equator. Black ellipses, circles, and squares are inserted to highlight differences in the regimes of interest (see text for details).} \label{fig:heightmaps_comp}
    \end{figure*}
    
\begin{figure*}
        \centering
        \includegraphics{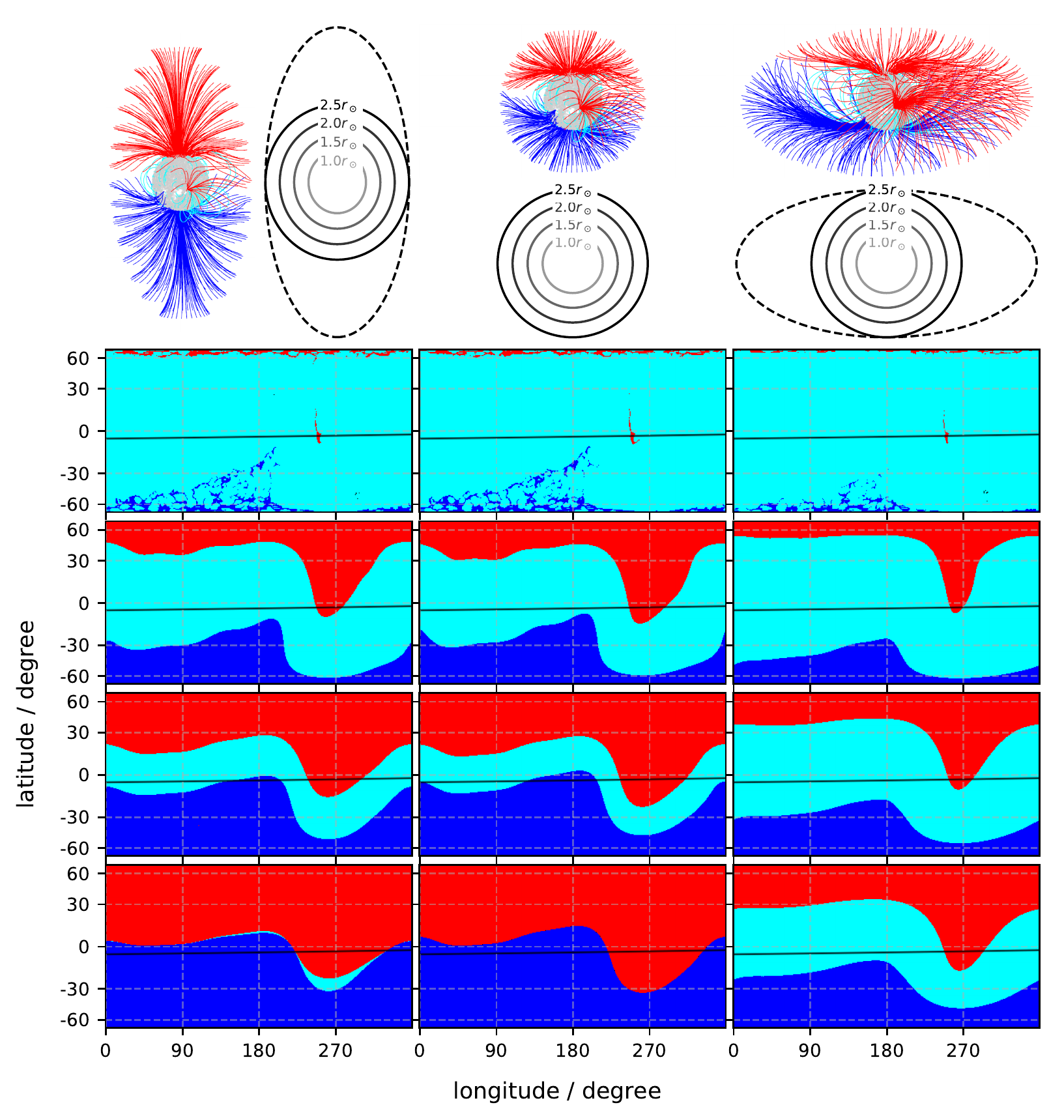}
        \caption{Same as in Fig. \ref{fig:heightmaps_comp}, but the height levels are spheres. The heights are again $1.0 R_\odot$, $1.5R_\odot$, $2.0R_\odot$, and $2.5R_\odot$. The dashed line in the first row depicts the source surface, which here is not the same as the uppermost height level examined. The uppermost levels exhibit open field structures in the ellipsoidal cases because the sphere of $2.5R_\odot$ is partially within the source surface and touches it at the equator (prolate case) or the poles (oblate case). \label{fig:heightmaps_world}}
    \end{figure*}

Ideally, an accurate algorithm for the computation of the solar magnetic field would accommodate arbitrary heights of the source surface at every longitudinal and latitudinal position. We propose a small step in this direction by implementing an ellipsoidal source surface controlled by a single parameter $A$. An oblate source surface in the PFSS model creates a magnetic field that reaches radial orientation closer above the poles compared to equatorial regions, as can be seen in the renderings in Figs. \ref{fig:heightmaps_comp} to \ref{fig:expansion_world}. Let $R_{ss}$ be the source surface height of the classical PFSS model. The actual source surface height $r_{ss}$ in our model is given by

\begin{align*}
    \mbox{oblate case:} && r_{ss,\mathrm{\tiny{equator}}} &= A\cdot R_{ss}\\
                        && r_{ss,\mathrm{\tiny{poles}}} &= R_{ss}\\
    ~\\
    \mbox{prolate case:}&& r_{ss,\mathrm{\tiny{equator}}} &= R_{ss}\\
                        && r_{ss,\mathrm{\tiny{poles}}} &= A\cdot R_{ss}.
\end{align*}

As the Sun still needs to be approximated as a sphere, we cannot simply employ a homogeneous elliptical grid. Hence, we created a grid that exhibits spherical symmetry at the photosphere while incrementally deforming higher grid shells to the ellipsoidal shape. A brief explanation of the grid is summarized in Appendix \ref{sec:algo}.

The elliptical implementation works in the same manner as the spherical one by repeatedly solving Laplace's equation at all grid points until the accuracy threshold $p$ is surpassed at all grid points (see Eq. \ref{equ:termination}). 

Magnetic field lines are oriented perpendicular to the source surface, which defines an isopotential. In the spherical case, this means field lines are already oriented radially. To obtain the radial orientation of the magnetic field lines in the elliptical model, a spherical surface can be positioned above the source surface, and some form of interpolation technique, for example  employing splines, can be applied to ``bend'' the magnetic field lines into the radial orientation.

\subsection{Parameters for the grid solver and treatment of input magnetograms\label{sec:datatreatment}}

For all model evaluations we employ a source surface at a heliocentric height of $2.5R_\odot$ (minor half-axis in the elliptical version).
As input data, we used synoptic magnetograms from the Wilcox Solar Observatory for testing purposes and comparisons with data products published by \citet{Hoeksema2020}. Because the resolution of these maps is low, no image processing needs to be applied. For high-resolution grid tests and for the plots presented in this work we employed synoptic magnetograms from the Michelson Doppler Imager (MDI) on board the Solar and Heliospheric Observatory (SOHO) \citep{Scherrer1995a}.
The high-resolution synoptic magnetic maps produced by MDI are scaled down to 87x175 pixels using a Lancosz filter. In a second step these magnetograms are corrected for the monopole offset which is introduced by small changes of the photospheric magnetic field during the data acquisition period of about 27 days and for data gaps near the poles. 

A well-known difficulty of numerical models is the trade-off between high accuracy and computational demand. After systematically testing and comparing results of several distributions and densities of grid points we chose our numerical grid to have 35x87x175 grid points in radial, meridional, and zonal directions. Radial spacing increases geometrically from the photosphere up to the source surface with a geometric factor $q \approx 3.3\%$. The radial position of grid points on a height level $i$ ($i \in \{1,\dots , N_r-1\}$) in the spherical grid is $r_i = r_{i-1} + (r_{1}-R_{\odot})\cdot q^{i-1}$, where $N_r=35$ is the number of height levels in the radial direction and $r_{1}-R_{\odot}\approx 18.7$~Mm. For the elliptic case, this height dependence is distorted according to the stretching treatment presented in Appendix \ref{sec:algo}.
This geometric increase in radial grid point spacing allows  higher accuracy of the solver near the photosphere where magnetic gradients are strongest, while reducing the computational footprint in the higher regions where a lower accuracy is sufficient for acceptable results.

Polar boundaries are introduced by computing virtual values at the poles as being the average value of all northernmost and southernmost grid points at that specific height. The virtual polar points serve as the outermost neighbor for all adjacent grid points that make up its average. This decreases computation time as information of the solution is allowed to directly travel over the poles rather than only in the zonal direction.

As termination criterion (Eq. \ref{equ:termination}) we used $p=0.01$ which means the value change from one time step to the next is below 1\% at all grid points. Magnetic field line tracking is done by employing an adaptive Runge-Kutta-Fehlberg method of fourth order (RKF45), tri-linearly interpolating between computational grid points.

The spherical PFSS model is a special case of the elliptical PFSS model with ellipticity $A=1.0$. The elliptic solver gives the same values at all grid points within floating-point rounding accuracy for this special case compared to the purely spherical solver.

Our implementation has been written in C/C++ and employs the CUDA framework (version 9.2) by NVIDIA. The time-stepping solution process is quite basic and can be sped up significantly by employing more sophisticated solution processes. However, the computation time for one Carrington rotation with the resolution and accuracy presented here takes between 10 and 20 minutes on the NVIDIA GTX Titan which was released in 2013. For the studies presented here, the computation time is sufficiently fast.

\section{\label{sec:results}Results}

Without answering the question of which parameter for the ellipticity gives the most realistic coronal magnetic field configuration (if any), we  illustrate here the qualitative differences between the spherical and elliptical PFSS models. We chose Carrington rotation 2066 during the minimum between solar activity cycles 23 and 24 in early 2008 to illustrate the differences the model parameters incur. There are only a few and weak CMEs registered for this Carrington rotation \citep{Yashiro}, and a few coronal holes and active regions \citep{Barra2009}.

Figures \ref{fig:heightmaps_comp} and \ref{fig:heightmaps_world} depict the solar coronal magnetic field configuration. The figure consists of a three-dimensional rendering of the magnetic field configuration as seen from the vernal equinox in the first row. The other rows show the magnetic field polarities at different heights between the photosphere and the source surface. For illustrative purposes only, we chose a strong ellipticity ($A=2$) of the source surface which we assume is higher than a realistic configuration would exhibit.

The three-dimensional renderings consist of two magnetic field line mappings from the source surface down to the photosphere and vice versa. At the specific height (photosphere, source surface), an equidistant two-dimensional grid is spanned in sin(latitude) and longitude with  $-14.5/15.0 \leq \sin{\left(\mbox{latitude}\right)} \leq +14.5/15.0$ and $0\leq\mbox{longitude}~<~2\pi$. Each of these grid points constitutes the starting point of a magnetic field line, which is traced throughout the computational domain. Blue magnetic field lines have a positive sign pointing outwards, while red lines have a negative sign pointing inwards. Closed field lines originating and ending on the photosphere are in cyan. In the renderings, $15 \times 30$ magnetic field lines are illustrated starting on the source surface and the same number starting on the photosphere.

Similarly, rows 2 to 5 are cuts of magnetic field line mappings originating at intermediate heights but with a higher resolution of $200 \times 400$ field lines, corresponding to $200 \times 400$ pixels. The format is similar to synoptic Carrington maps. The uppermost and lowermost pixels again are positioned at $\sin(\mbox{latitude}_{\mbox{\tiny{max}}})=+14.5/15.0$ and $\sin(\mbox{latitude}_{\mbox{\tiny{min}}})=-14.5/15.0$, respectively. Each pixel represents the polarity of the magnetic field line at that pixel center position with the same color scheme as the field lines in the renderings. Due to the sine-latitude spacing pixels near the poles cover a larger area than pixels at the equator, effectively reducing resolution at high latitudes.

In Fig. \ref{fig:heightmaps_comp} the intermediate maps correspond to ellipsoidal surfaces which grow in ellipticity, as does the underlying computational grid when approaching the source surface from below.
Figure \ref{fig:heightmaps_world} shows intermediate height maps originating on spheres  between the two computational boundaries. These represent height levels of $1.0R_\odot$, $1.5R_\odot$, $2.0R_\odot$, and $2.5R_\odot$ heliocentric distance. We note that the middle column for the spherical model is the same in both figures.

Figure \ref{fig:heightmaps_comp}  considers global differences among the three presented models. Each height level depicts the magnetic polarity structure at the same relative distance between the boundaries, which correspond to different absolute heliocentric heights for each column. Figure \ref{fig:heightmaps_world} helps to visualize local differences. Each pixel represents the same physical position in all three columns.
While  most structures are present in every  model, their general  appearance can vary greatly. In Figs. \ref{fig:heightmaps_comp} and \ref{fig:heightmaps_world} the pronounced current sheet warp (rectangles) at longitude $270^\circ$, for example, is less pronounced in the prolate case and extends to lower latitudes in the oblate case compared to the spherical reference model.
Another example is the open positive structure slightly south of the equator at longitude $180^\circ$ (ellipses and circles) which is present in the spherical and prolate models, but  missing in the oblate version.
Also, closed field structures extend higher in the computational domain in the oblate case, as can be easily seen in the third row of Fig. \ref{fig:heightmaps_comp}.
     \begin{figure*}
        \centering
        \includegraphics{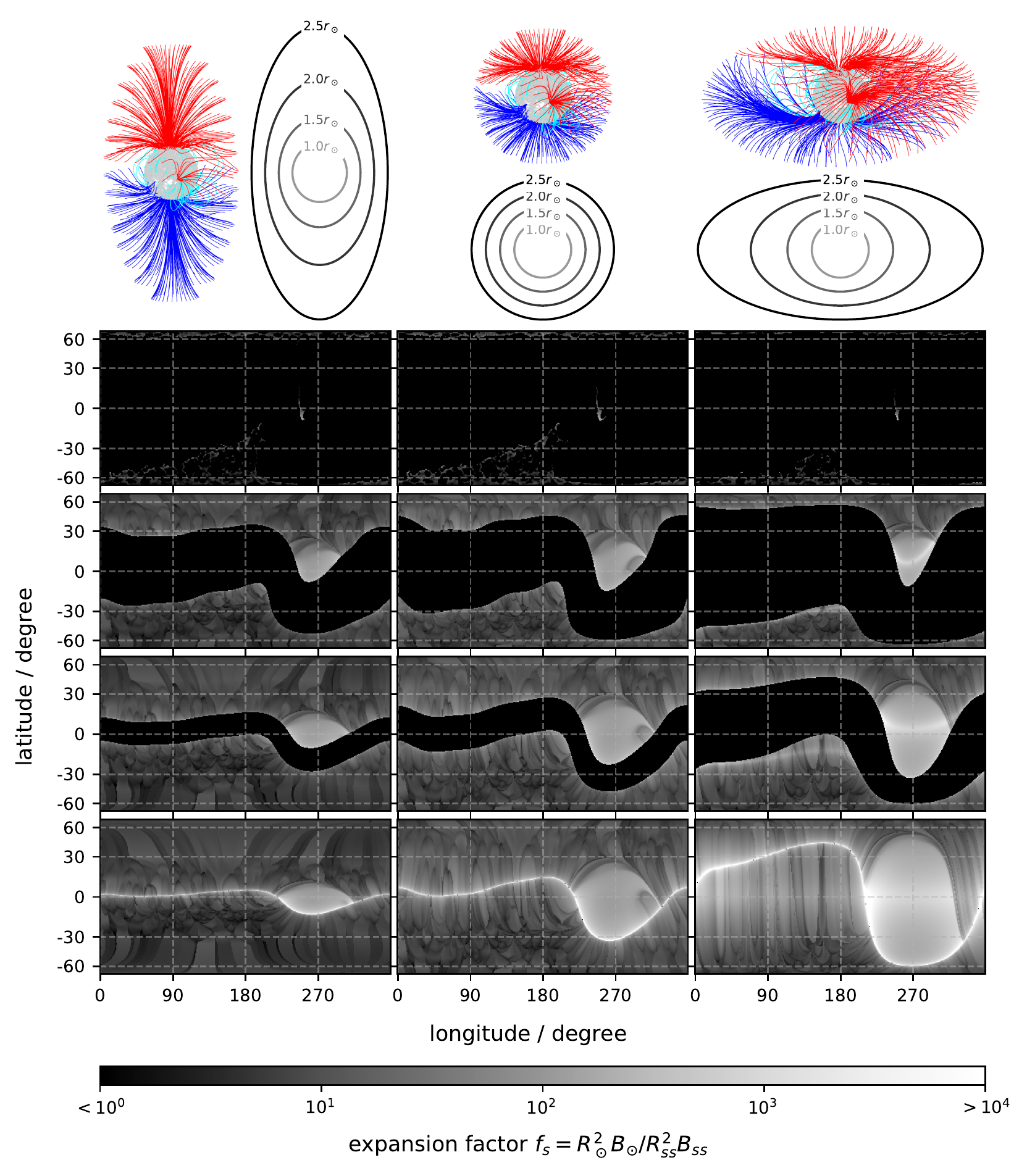}
        \caption{Height levels show the expansion factor. Each pixel again corresponds to one field line, and the color represents its expansion factor between photosphere and source surface.\label{fig:expansion}}
    \end{figure*}
    
     \begin{figure*}
     \centering
        \includegraphics{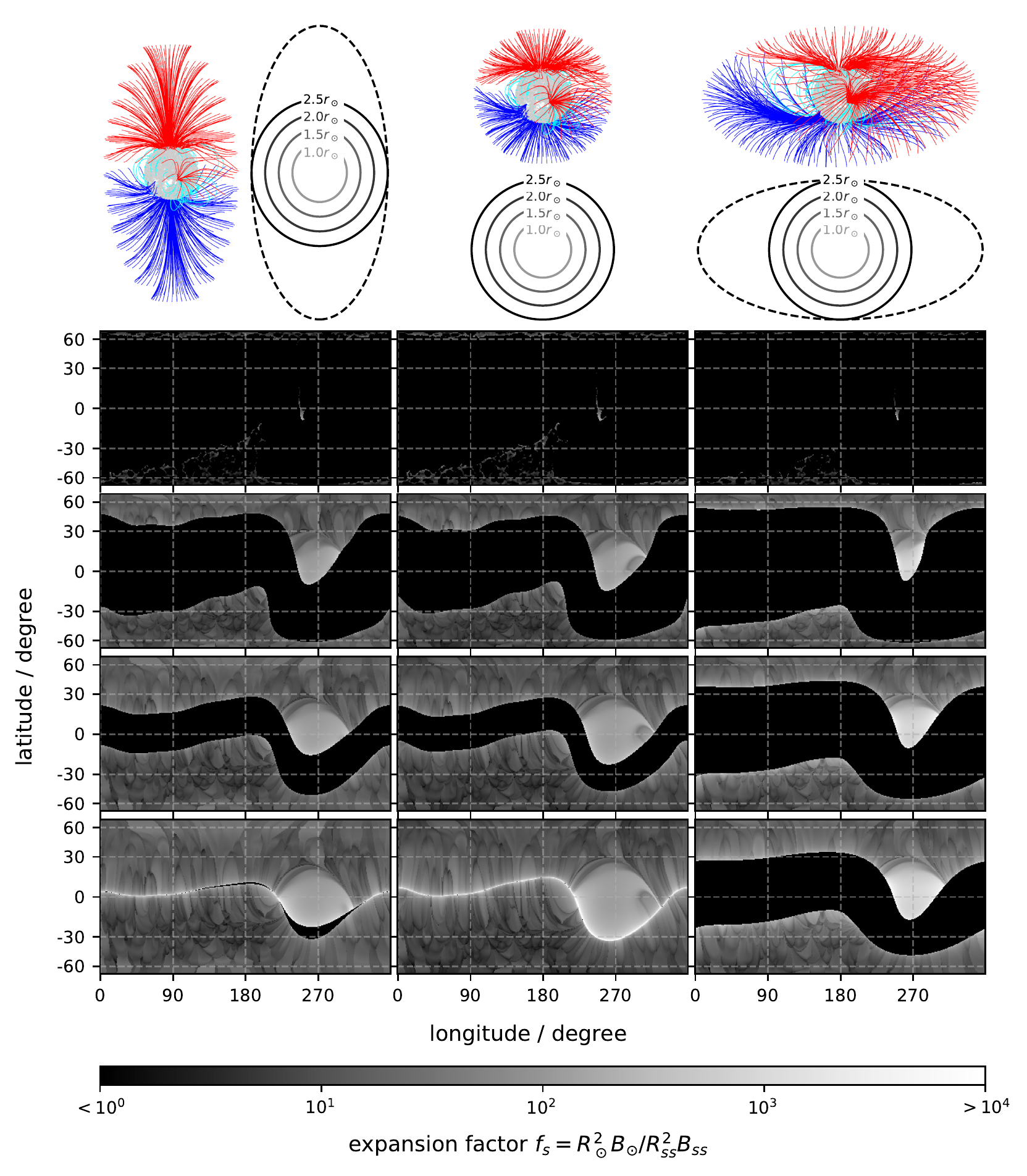}
        \caption{Same as Fig. \ref{fig:expansion}, but for height levels as spheres.\label{fig:expansion_world}}
    \end{figure*}
    
Another characteristic that can be computed by only considering magnetic field data is the flux-tube expansion factor $f_s = B_\odot R_\odot^2 / B_{ss} r_{ss}^2$ \citep{Wang1990}, where $B_\odot$ is the magnetic flux density at the photosphere, $B_{ss}$ is the magnetic flux density at the source surface, $R_{\odot}$ is the heliocentric distance of the photosphere, and $r_{ss}$ is the heliocentric distance of the source surface.

The flux-tube expansion factor is inversely related to solar wind speed and offers a means of comparing model prediction to in situ spacecraft data (ibid). In Figs. \ref{fig:expansion} and \ref{fig:expansion_world}, we illustrate the expansion factor for the three PFSS models. The format is similar to Figs. \ref{fig:heightmaps_comp} and \ref{fig:heightmaps_world}, but here each pixel is color-coded with the expansion factor for that specific field line. The oblate PFSS model exhibits higher expansion factors at lower latitudes than the other two models. This suggests lower solar wind speeds in the oblate model, which should be verifiable by analyzing spacecraft data.

\subsection{\label{sec:shccomp}Comparison of the classic implementation to the grid solver}

\begin{figure*}
        \includegraphics{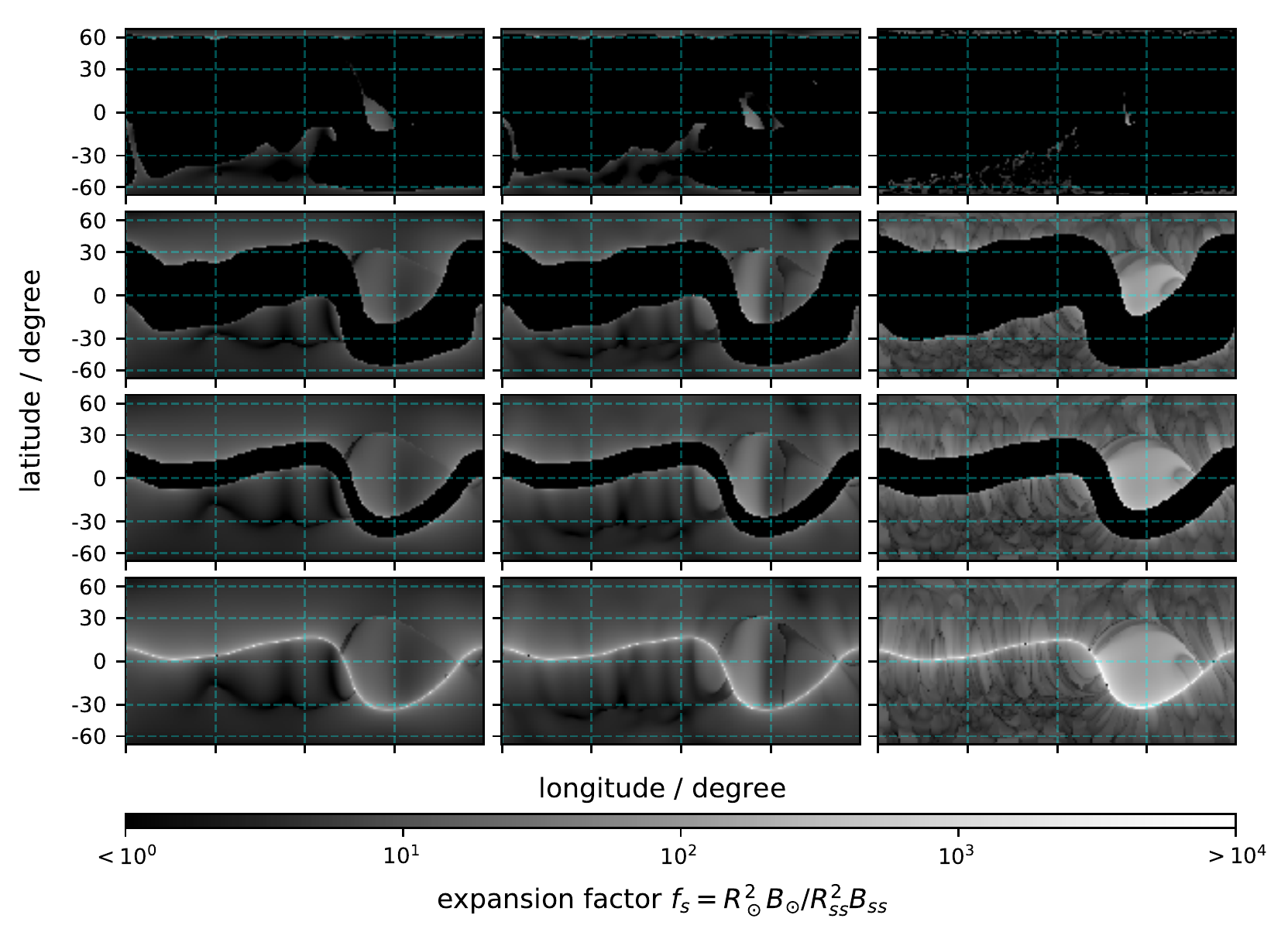}
        \caption{Comparison of expansion height maps for the classical PFSS implementation using spherical harmonic coefficients with order 9 in the left column, with order 20 in the middle column, and our grid solver in the right column. Again Carrington rotation 2066 is depicted. All models employ a spherical source surface at $2.5 R_\odot$ heliocentric distance. The height levels in rows 1 to 4 are spheres at heliocentric distances of $1.0 R_\odot$, $1.5 R_\odot$, $2.0 R_\odot$, and $2.5 R_\odot$.\label{fig:heightmaps_shc}}
    \end{figure*}

The expansion factor also illustrates some differences in resolution between the classic implementation employing spherical harmonic coefficients and our grid solver. Figure \ref{fig:heightmaps_shc} depicts the expansion factor height levels for the SHC implementation of orders 9 and 20 as well as our implementation. Unsurprisingly, the middle column, which corresponds to a maximum principle order 20, exhibits more detail than the left column, which corresponds to  the classical approach of order 9. Our grid implementation in the right column offers even more detail at the cost of longer computation times. While the classical SHC implementation of order 9 takes less than 1 minute using a single thread on the CPU (Intel Xeon E5-1650), our grid solver takes about 15 minutes employing the massively parallel architecture of a GPU (GTX Titan).

\section{\label{sec:conclusion}Discussion and conclusions}

The alteration to the PFSS model to incorporate elliptical source surfaces allows us to tweak the magnetic field computations without  needing to employ the full set of MHD assumptions and computational complexity. Only one parameter has been added that needs to be determined and evaluated. In this regard, it is comparable with other improvements of the PFSS model like CSSS, which includes two additional parameters, namely the height of the cusp surface and a length scale of the assumed horizontal electric currents in the corona \citep{Zhao1995}. 

The CSSS model still relies on a spherical symmetry which is a strict constraint for a physical system that we know deviates from this type of symmetry. It is known that the ``true'' source surface, if it even exists, is most probably not elliptical either \citep[see, e.g.,][]{Cohen2015,Schulz1978,Riley2006,Panasenco2020}. It remains to be seen whether the elliptical PFSS approximates the true magnetic field configuration better or worse than the other improvements that have been developed by the community in the past decades. It might also be useful to combine the elliptical source surface with these models, thereby merging possible advantages either model has over the original spherical PFSS model.

The finite difference solver allows  more complex shapes to act as source surfaces. For example, the polar indentations found by \cite{Riley2006} may be modeled using the same techniques presented here. 

Phenomenologically, there are a few differences between the three grid models presented in this work. In the oblate elliptical model, magnetic field lines tend to bend towards the poles, while in the prolate model they bend away from the poles (see the renderings in Figs. \ref{fig:heightmaps_comp} to \ref{fig:expansion_world}). Hence, particles traveling along these field lines will have slightly different trajectories. In the oblate case, closed magnetic field structures extend higher relative to the source surface compared to the other two models.

One way to evaluate PFSS models is to compare in situ spacecraft measurements of the heliospheric magnetic field with predictions of the PFSS model. After measuring the solar wind speed at the spacecraft, the likely footpoint of the solar wind plasma package on the source surface can be computed by tracing the Parker spiral. The magnetic polarity at the footpoint, as computed by the PFSS implementation, can then be compared to the magnetic field measured at the spacecraft. While there are phenomena above the source surface that might alter the magnetic field orientation as well as inaccuracies in tracing the correct Parker spiral, for example due to solar wind acceleration processes, a better magnetic model of the corona is expected to increase the agreement of spacecraft data with predicted model data.

Going one step further it is possible to track the source surface footpoint of the spacecraft down to the photosphere along the magnetic field lines computed by the model. The photospheric footpoint positions can then be compared to the appropriate positions in extreme ultraviolet (EUV) images, which originate close to the photosphere. It is generally assumed that open magnetic flux structures map to darker regions in EUV images \citep{Huang2019}. A better magnetic field model should therefore map more often to dark regions in EUV maps than do worse models.

Evaluating predictions of the PFSS models utilizing spacecraft data is complicated due to the distances between spacecraft (typically at heliocentric distances of about 1AU) and the computational domain (below a few $R_\odot$). Connecting spacecraft positions to the source surface introduces errors increasing with distance. Fortunately, two recent missions might help this endeavor. The Parker Solar Probe \citep{Fox2016} is already collecting data and  Solar Orbiter \citep{Muller2013} was launched in February 2020. The Parker Solar Probe is scheduled to reach a perihelion heliocentric distance of less  than $10 R_\odot$, while the Solar Orbiter will go as close as $60 R_\odot$.  Solar Orbiter will have a higher ecliptic inclination allowing for measurements closer to the solar poles.

Having instruments close to the computational domain of the PFSS model will help to evaluate the alterations made in this work significantly. Data from both Solar Orbiter as well as the Parker Solar Probe are expected to be available during the 2020s.

\cite{Badman2019} already employed a spherical PFSS model and found that a lower source surface between $1.3 R_\odot$ and $1.5 R_\odot$ matched the observations better than the traditional source surface height of $2.5 R_\odot$. They also pointed out that this exceptionally low source surface might compensate for washed out small-scale structures due to traditional modeling parameters, which the model alterations presented here might help to alleviate.

\begin{acknowledgements}
      Wilcox Solar Observatory data used in this study was obtained via the web site http://wso.stanford.edu at 2019:10:31\_09:49:01 PDT courtesy of J.T. Hoeksema.\\
      
      This work uses data from the Michelson Doppler Imager (MDI) onboard the Solar and Heliospheric Observatory (SOHO). SOHO is a project of international cooperation between ESA and NASA to study the Sun, from its deep core to the outer corona, and the solar wind. 
\end{acknowledgements}

\appendix
\section{\label{sec:algo}Mathematical framework of the elliptical grid solver}
The framework of the employed grid is not trivial, in particular the transition from the spherical configuration at the base to elliptical at the source surface. Here we briefly describe the framework employed for this work. The derivation follows the procedure laid out by \citet{Piercey2007} for non-orthogonal curvilinear coordinate systems. 

To simplify the mathematical description, we distinguish between two domains. The first domain is a basic spherical coordinate system on which the algorithm performs its operations. This domain is called the computational domain. By stretching this domain along one or two axes we obtain the physical domain, which corresponds to real-world coordinates that we are interested in. By employing this two-domain approach we circumvent the need for complicated derivations of the mathematical framework by hiding it in a very simple transformation from the computational to the physical domain.
Stretching the computational grid along one axis (z) produces a prolate ellipsoidal source surface. Stretching the grid along two axes (x and y) produces an oblate ellipsoid. In the following, we concentrate on the oblate case. The prolate version is obtained in an analogous manner.

    \begin{figure}
        \includegraphics{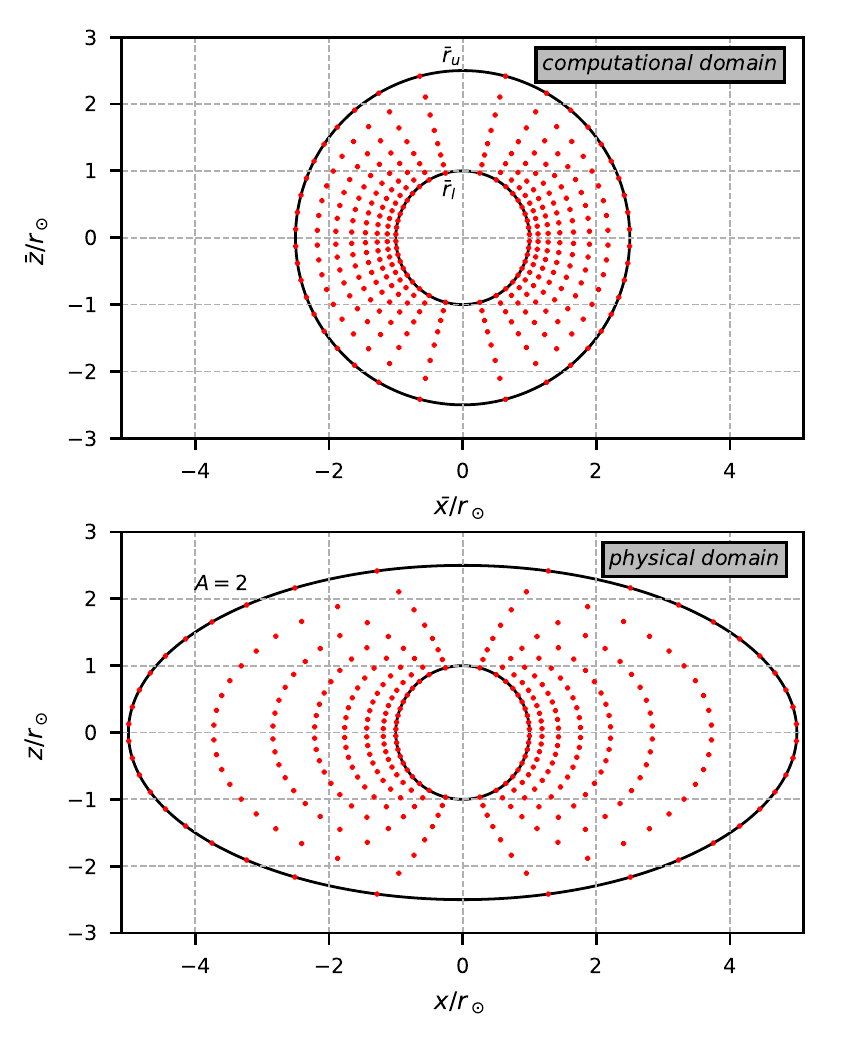}
        \caption{Cut through the $\bar{x}\bar{y}/xy$-plane of the computational and physical domain for an oblate ellipsoidal source surface. The number of grid points shown is considerably reduced to improve clarity. The presented grid has an ellipticity of $A=2$ at the source surface, which results in a source surface height of $2.5R_\odot$ over the poles and of $5.0R_\odot$ over the equator.\label{fig:comp_phys_domain}}
    \end{figure}

Figure \ref{fig:comp_phys_domain} shows cuts through the two domains. Coordinates in the computational domain are denoted by bars above the symbols ($\bar{x}$, $\bar{y}$, $\bar{z}$, $\bar{r}$, $\bar{\theta}$, $\bar{\phi}$), whereas the same symbols without bars are used for the physical domain ($x$, $y$, $z$, $r$, $\theta$, $\phi$). They are in two groups: $x$, $\bar{x}$, $y$, $\bar{y}$, $z$ and $\bar{z}$ are the well-known cartesian coordinates and $r$, $\bar{r}$ $\theta$, $\bar{\theta}$, $\phi$, and $\bar{\phi}$ are the spherical coordinates in their respective domains.
The stretching of the computational grid is performed by an analytic stretching function along the $\bar{x}$- and $\bar{y}$-axes according to

\begin{align*}
    x &= {a}\bar{x} = {a_A}(\bar{r}) \bar{r}\sin\bar{\theta}\cos\bar{\phi} \\
    y &= {a}\bar{y} = {a_A}(\bar{r}) \bar{r}\sin\bar{\theta}\sin\bar{\phi} \\
    z &= \bar{z} = \bar{r}\cos\bar{\theta}
\end{align*}

As the stretching function, we chose

\begin{equation*}
    {a_A}(\bar{r}) = 1 + \frac{A-1}{\bar{r_u}^2-\bar{r_l}^2}\left(\bar{r}^2-\bar{r_l}^2\right) =2\alpha + a_s\bar{r}^2,
\end{equation*}
where $A$ is the ellipticity parameter at the source surface; $\bar{r_u}$ and $\bar{r_l}$ are the radial positions of the upper and lower computational boundaries, respectively; $\alpha = (1-a_s \bar{r_l}^2)/2$; and $a_s = (A-1)/(\bar{r_u}^2-\bar{r_l}^2)$. 
We chose the squared dependence on radial distance in the computational domain to decrease the rate of change in the lower region where the algorithm requires a higher computational accuracy compared to the outer region near the source surface.

With these relations between computational and physical domain the gradient basis vectors of the physical coordinate system are 

\begin{align*}
    \vec{g}^r &= \frac{1}{a + r\cdot\sin^2{\theta}\cdot\frac{\partial a}{\partial r}}\left(\begin{array}{c}
        \sin{\theta}\cos{\phi}\\
            \sin{\theta}\sin{\phi}\\
            a\cos{\theta}\end{array}\right) \\
        \vec{g}^\theta &= \frac{1}{r\left(a+r\cdot\sin^2{\theta}\cdot\frac{\partial a}{\partial r}\right)}\left(\begin{array}{c}
            \cos{\theta}\cos{\phi} \\
            \cos{\theta}\sin{\phi} \\
            -\sin{\theta}\left(a+r\cdot\frac{\partial a}{\partial r}\right)\end{array}\right)\\
        \vec{g}^\phi &= \frac{1}{a\cdot r\cdot \sin{\theta}}\left(\begin{array}{c}
        -\sin{\phi}\\
        \cos{\phi}\\
        0\end{array}\right).
\end{align*}

With these vectors and a twice continuously differentiable scalar function $\Psi$, the Laplace operator in general curvilinear coordinates can be expressed as

\begin{align*}\label{equ:laplace}
        \nabla^2\Psi = \frac{1}{\sqrt{g}}&\sum_{i}\sum_{j}\frac{\partial}{\partial q^i}\left(\sqrt{g}g^{ij}\frac{\partial\Psi}{\partial q^j}\right),\\
\end{align*}
where $g=det(G)$ is the determinant of the metric coefficient matrix $G$ with entries $g^{ij} = \vec{g}^i\cdot\vec{g}^j$, and $i,j\in \left\{1,2,3\right\}$ corresponding to the coordinates $r$, $\theta$, and $\phi$, respectively. 

\bibliographystyle{aa}
\bibliography{bib}

\end{document}